\documentclass[twocolumn,showpacs,preprintnumbers,amsmath,amssymb,prl]{revtex4}
\usepackage{graphicx}
\usepackage{pifont}
\usepackage{graphicx}
\usepackage[mathscr]{eucal}
\newcommand{\D}{\displaystyle}

\begin{document}

\title{Cis-Regulatory Modules Drive Dynamic Patterns of a Multicellular System}

\author{Jiajun Zhang$^1$}
\author{Zhanjiang Yuan$^1$}
\author{Tianshou Zhou$^{2,1,}$}
\email{mcszhtsh@mail.sysu.edu.cn}

\affiliation{\\$^1$School of Mathematics and Computational Science,
Sun Yat-Sen University, Guangzhou 510275, China\\
$^2$State Key Laboratory of Biocontrol and Guangzhou Center for
Bioinformatics, School of Life Science, Sun Yat-Sen University,
Guangzhou 510275, China}

\date{\today}

\begin{abstract}
How intracellular and extracellular signals are integrated by
transcription factors is essential for understanding complex
cellular patterns at the population level. In this Letter, by using
a synthetic genetic oscillator coupled to a quorum-sensing
apparatus, we propose an experimentally feasible cis-regulatory
module (CRM) which performs four possible logic operations (ANDN,
ORN, NOR and NAND) of input signals. We show both numerically and
theoretically that these different CRMs drive fundamentally
different dynamic patterns, such as synchronization, clustering and
splay state.
\end{abstract}


\maketitle

Biological organisms possess an enormous repertoire of genetic
responses to ever-varying combinations of cellular and environmental
signals \cite{BookDavidson01,BookAlon06}. Such a repertoire is
typically encoded in complex regulatory networks, and affects
patterning, differentiation and growth. At the heart of these
networks are cis-regulatory modules (CRMs), which contain a cluster
of binding sites for transcription factors (TFs) and determine the
place and timing of gene action within the network. Both deciphering
the codes and elucidating the functions of CRMs involved in various
developmental processes are a major challenge in biology.

It has been shown that CRMs can perform an elaborate computation at
the individual gene level: the transcription rate of a gene depends
on the active concentration of each of inputs
\cite{Buchler03,PlosComputBiol06,Mangan03,Setty03,PlosBiol06}. On
the other hand, cells live in a complex environment and can sense
many different signals, in particular those from neighboring cells.
Therefore, at the multicell level CRMs need to integrate
intracellular and extracellular signals so as to coordinate gene
expression. Given that cells are frequently subject to chemical
signals from neighboring cells, it is worth studying the effect of
chemical communication on the dynamic patterns of multicellular
systems. Modeling studies, for example, have shown that a population
of repressilators coupled to quorum sensing can work as a
macroscopic genetic clocks \cite{Garcia-Ojalvo04}. In that study,
two input signals (i.e., two TFs) regulate a target gene
independently. TFs, however, are often integrated in a combinatorial
logic manner, and moreover such a combination may take different
forms \cite{Buchler03,PlosComputBiol06,Mangan03,Setty03,PlosBiol06}.
From views of evolutionism, CRMs are changeable, e.g.,
cis-regulatory mutations \cite{AWGregory07}. Such a mutation
constitutes an important part of the genetic basis for adaptation. A
naturally arising question is how the changes of CRMs affect
cellular patterns of populations of genetic oscillators. We address
this question by designing a multicellular network with a CRM
consisting of repressilators \cite{Repressilator} coupled to quorum
sensing \cite{Garcia-Ojalvo04,McMillen02,Fuqua96} in {\it
Escherichia coli}. In contrast to the previous studies
\cite{Garcia-Ojalvo04,McMillen02} that numerically showed that
coupled genetic oscillators can demonstrate synchronous behaviors,
we both numerically and theoretically show that different signal
integration (ANDN, ORN, NOR, NAND type of responses) leads to
fundamentally different properties, such as synchronization,
clustering, and splay state. Our results indicate that the CRM has a
significant influence on the mode of cellular patterns.

\begin{figure*}[htb]
\begin{center}
\includegraphics [width=16cm] {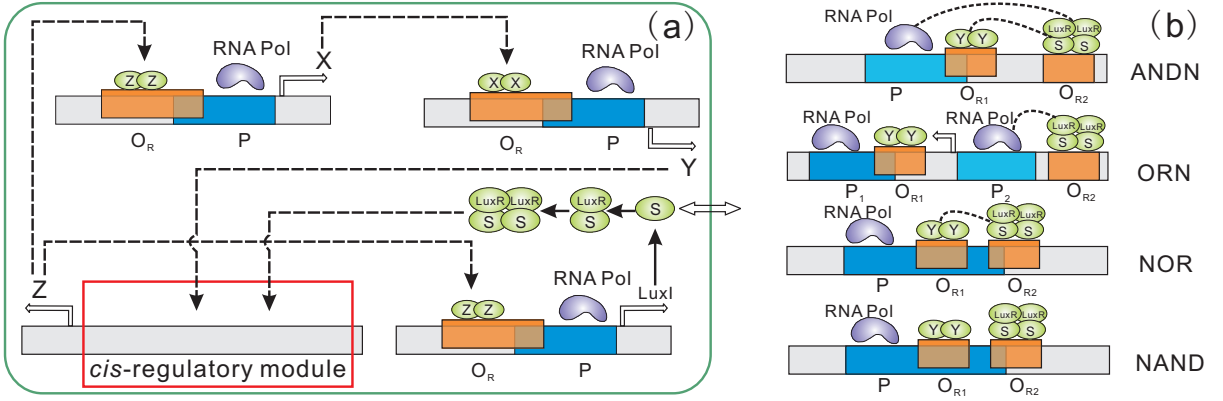}
\caption{(color). (a) The schematic diagram of a multicellular
system with a cis-regulatory module (CRM). Three transcriptional
repressors (X, Y, Z) inhibit one another in a cyclic way. The gene
{\it luxI} from the LuxI/LuxR module first synthesizes a small
molecule S. Both S and LuxR then form a hetero-tetramer complex. The
dimer of Y and the complex co-regulate the target gene, thus
carrying out the function of a CRM (symbolled by the empty box). The
right bidirectional arrow indicates that S can freely diffuse
through the cellular membrane. (b) Four cis-regulatory constructs
for implementations of four different logic functions. From top to
bottom is ANDN, ORN, NOR and NAND, respectively. In ANDN, the
activator S is unable to act if the repressor Y is bound to the
promoter; In ORN, the CRM is constructed using a weak promoter and a
strong promoter, where the activator S is able to act if the
repressor Y is bound to the promoter; In NOR, two repressors S and Y
produce the full repression cooperatively; In NAND, the promoter is
regulated exclusively by two repressors. In (a) and (b), P and
O$_{\rm R}$ denote the promoter (dark blue box for the strong
promoter and light blue box for the weak promoter) and the operator
site (jacinth box) respectively, and the RNA Pol represents RNA
polymerase. We use offset and overlapping boxes to indicate the
mutual repression and the dashed lines to indicate the cooperative
interaction.}\label{f1}
\end{center}
\vspace*{-0.5cm}
\end{figure*}

A multicellular network under investigation is schematically shown
in Fig. 1(a). In such a network, the signaling molecule (S) carries
out the information exchange between cells and regulates the
expression of a target gene through a CRM. The S and the TF (Y)
first bind to specific DNA sequences of the CRM, and then
co-regulate the expression of the gene in a combinatorial scheme. In
theory, this type of CRM can perform eight different cis-regulatory
input functions (CRIFs) \cite{PlosComputBiol06}, but limited by the
cyclic repression structure of repressilator, we have only four
types of CRIFs: ANDN, ORN, NOR and NAND (see Ref. \cite{Supp} for
exact explanations). Figure 1 (b) gives the detailed regulation
scheme of every CRM.

Based on the biochemical reactions given in Table 1 and defining the
rescaled concentrations as our dynamical variables, the
dimensionless equations of intracellular dynamics are described as
\begin{eqnarray} \label{eq1}
\frac{d{\bf X}_i}{dt}&=& {\bf F}({\bf X}_i,S_i)\,,\\
\frac{dS_i}{dt}&=& E({\bf X}_i,S_i)+\eta(S_e-S_i)\,,
\end{eqnarray}
where subscript $i$ represents cell $i$ ($=1,2,\cdots,N$), and ${\bf
X}_i=(x_i,\,y_i,\,z_i,\,X_i,\,Y_i,\,Z_i)^{\rm T}$ with $x_i$, $y_i$
and $z_i$ standing for three mRNA concentrations, and $X_i$, $Y_i$
and $Z_i$ for three protein concentrations. $S_i$ represents the
concentration of the signaling molecule inside the $i$th cell
whereas $S_e$ does the concentration of the signal in the
extracellular environment. Because of the fast diffusion of the
extracellular signal compared to the repressilator period, $S_e$ can
be assumed to be in the quasi-steady state, leading to
$S_e=\frac{Q}{N}\sum_{i=1}^N S_i$, where the parameter $Q$ depends
on the cell density in a nonlinear way \cite{Garcia-Ojalvo04}. ${\bf
F}=(F_1,F_2,\ldots,F_6)^{\rm T}$ with $F_1=\frac{\alpha}{1+Z^n}-x$,
$F_2=\frac{\alpha}{1+X^n}-y$, $F_4=\beta(x-X)$, $F_5=\beta(y-Y)$,
$F_6=\beta(z-Z)$, $E=\gamma X-\delta S$, and
\begin{eqnarray}
F_3={\rm CRIF}- z\,,
\end{eqnarray}
where we omit subscript $i$ for convenience, and the core function
CRIF corresponding to ANDN, ORN, NOR and NAND respectively is listed
in the first part of Table 1. The detailed derivation of CRIFs and
${\bf F}$ is put in Ref. \cite{Supp}. Throughout this Letter, all
parameters except for $Q$ are set as $\alpha=204$, $\beta=1$,
$\gamma=0.01$, $\delta=1$, $n=2$, $\eta=2$, $\mu=51$, $\nu=204$,
$\lambda=1$, which come from experimentally-reasonable settings
\cite{Supp}. Since the numerical results do not depend qualitatively
on the cell number, we set $N=120$.
\begin{table}
\caption{Biochemical reactions and cis-regulatory input functions
(CRIFs). See Ref. \cite{Supp} for the derivation of CRIFs,
experimental values of parameters (including $\mu$, $\nu$ and
$\lambda$ that depend on reaction rates), and meanings of all used
symbols.}\vspace{-0.4cm} \label{t1}
\[ \begin{array}{cccccc}\hline\hline
\multicolumn{2}{c}{\mbox{Logic Function}} & \multicolumn{2}{c}{\mbox{\hspace{0.5cm}CRIF}} & \multicolumn{2}{c}{\hspace{0.5cm}\mbox{Reactions}}\\
\hline
\multicolumn{2}{c}{\rule[0cm]{0mm}{0.583cm}\mbox{ANDN}} & \multicolumn{2}{c}{\hspace{0.5cm}\D\frac{\mu S^2}{1 + S^2 + Y^2 + \lambda S^2Y^2}} & \multicolumn{2}{c}{\hspace{0.5cm}\mbox{\ding {172}\ding {173}\ding {174}\ding{176}}}\\[2ex]
\multicolumn{2}{c}{\mbox{ORN}} & \multicolumn{2}{c}{\hspace{0.5cm}\D \frac{\mu S^2+\nu}{1 + S^2 + Y^2}} & \multicolumn{2}{c}{\hspace{0.5cm}\mbox{\ding {172}\ding {174}\ding {175}\ding {176}}}\\[2ex]
\multicolumn{2}{c}{\mbox{NOR}} & \multicolumn{2}{c}{\hspace{0.5cm}\D \frac{\nu}{1 + S^2 + Y^2 + \lambda S^2Y^2}} & \multicolumn{2}{c}{\hspace{0.5cm}\mbox{\ding {172}\ding {173}\ding {174}\ding {175}}}\\[2ex]
\multicolumn{2}{c}{\mbox{NAND}} & \multicolumn{2}{c}{\hspace{0.5cm}\D \frac{\nu}{1 + S^2 + Y^2}} & \multicolumn{2}{c}{\hspace{0.5cm}\mbox{\ding {172}\ding {174}\ding {175}}}\\[2ex]\hline
\multicolumn{3}{l}{\mbox{\hspace{0.45cm}Fast Reactions}} & \multicolumn{3}{l}{\mbox{\hspace{0.7cm}Slow Reactions}}\\
\hline \multicolumn{3}{l}{\left. \begin{array}{l}
 2{\rm X} \overset{K_1}{\underset{}{\rightleftharpoons}} {\rm X}_2;2{\rm Y} \overset{K_2}{\underset{}{\rightleftharpoons}} {\rm Y}_2\\[-0.0ex]
 2{\rm Z} \overset{K_3}{\underset{}{\rightleftharpoons}} {\rm Z}_2;2{\rm C} \overset{K_4}{\underset{}{\rightleftharpoons}} {\rm C}_2\\[-1.1ex]
 {\rm S}+{\rm LuxR} \overset{K_5}{\underset{}{\rightleftharpoons}} {\rm C}\\[-1.1ex]
 {\rm D}^Y+{\rm X}_2 \overset{K_6}{\underset{}{\rightleftharpoons}} {\rm D}_X^Y\\[-1.1ex]
 {\rm D}^Z+{\rm Y}_2 \overset{K_7}{\underset{}{\rightleftharpoons}} {\rm D}_Y^Z\\[-1.1ex]
 {\rm D}^X+{\rm Z}_2 \overset{K_8}{\underset{}{\rightleftharpoons}} {\rm D}_Z^X\\[-1.1ex]
 {\rm D}^Z+{\rm C}_2 \overset{K_9}{\underset{}{\rightleftharpoons}} {\rm D}_C^Z\\[-1.1ex]
 {\rm D}^L+{\rm Z}_2 \overset{K_{10}}{\underset{}{\rightleftharpoons}} {\rm D}_Z^L
 \end{array}\right\}\mbox{\ding {172}}}
 & \multicolumn{3}{l}{\left. \begin{array}{l}
 {\rm D}^X \overset{k_X}{\rightharpoonup} {\rm D}^X + {\rm mRNA}_X\\[.3ex]
 {\rm D}^Y \overset{k_Y}{\rightharpoonup} {\rm D}^Y + {\rm mRNA}_Y\\[.3ex]
 {\rm D}^L \overset{k_L}{\rightharpoonup} {\rm D}^L + {\rm mRNA}_L\\[.3ex]
 {\rm L} \overset{c}{\rightharpoonup}{\rm L}+{\rm S};{\rm S} \overset{d_S}{\rightharpoonup} \varnothing\\[.3ex]
 {\rm mRNA}_I \overset{t_I}{\rightharpoonup} {\rm mRNA}_I+I\\[.3ex]
 {\rm mRNA}_I \overset{e_I}{\rightharpoonup} \varnothing\\[.3ex]
 {\rm I} \overset{d_I}{\rightharpoonup} \varnothing \\[.3ex]
 \small{\mbox{(I = X, Y, Z, L)}}
 \end{array}\right\}\mbox{\ding {174}}}\\[-0.7ex]
\multicolumn{3}{l}{\begin{array}{l}
 {\rm D}_C^Z+{\rm Y}_2 \overset{K_{11}}{\underset{}{\rightleftharpoons}} {\rm D}_{CY}^Z\\[-1ex]
 {\rm D}_Y^Z+{\rm C}_2 \overset{K_{12}}{\underset{}{\rightleftharpoons}} {\rm D}_{YC}^Z
 \end{array}\hspace{0.25cm}\Bigg\}\hspace{0.1cm}\mbox{\ding {173}}}
 & \multicolumn{3}{l}{\begin{array}{l}
 {\rm D}^Z \overset{k_Z}{\rightharpoonup} {\rm D}^Z+{\rm mRNA}_Z \hspace{0.53cm}\big\}\hspace{0.09cm}\mbox{\ding {175}}\\
 {\rm D}_C^Z \overset{fk_Z}{\rightharpoonup} {\rm D}_C^Z+{\rm mRNA}_Z\hspace{0.37cm}\big\}\hspace{0.09cm}\mbox{\ding
 {176}}^{\S}
 \end{array}}\\[-0.0ex]
\hline\hline \multicolumn{6}{l}{\scriptsize{\mbox{$^{\S}$ In ORN,
$k_Z$ in \ding{176} differs from that in \ding{175} due to different
promotors.}}}
\end{array}  \]
\vspace{-0.5cm}
\end{table}

\begin{figure}[t]
\begin{center}
\includegraphics [width=\hsize]{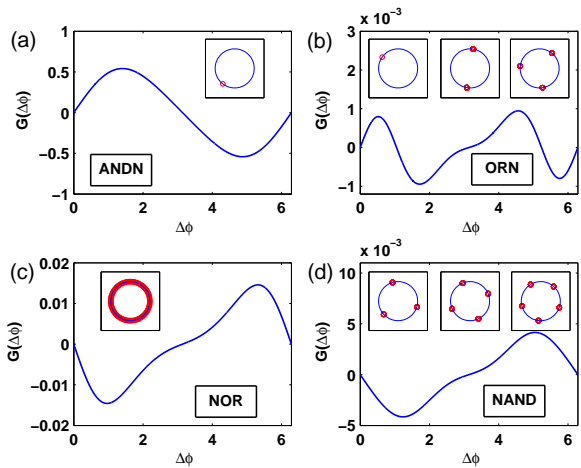}
\caption{(color). Different cis-regulatory modules drive different
cellular patterns. Insets display instantaneous distributions of
phases of the oscillators for a fixed $Q=0.5$: (a) 1-cluster state
(complete synchronization) for ANDN; (b) 1-, 2- and 3-cluster states
for ORN; (c) splay state for NOR; (d) 3-, 4- and 5-cluster states
for NAND (the corresponding time courses of all clustering figures
are put in the supporting material Ref. \cite{Supp}). The function
$G(\Delta\phi)$ determines the coupling mode: attractive coupling
for ANDN and ORN (due to $G'(0)>0$) and repulsive coupling for NOR
and NAND (due to $G'(0)<0$). Here, the different clusterings arise
from different initial phases.}\label{f1}
\end{center}
\vspace{-0.5cm}
\end{figure}

We are interested in the influence of four possible CRIFs on
cellular patterns. The results shown in the insets of Fig. 2
indicate that these different CRMs drive fundamentally different
dynamic patterns. Specifically, in the case of ANDN, for arbitrarily
chosen initial conditions we observe complete synchronization
(1-cluster) only, similar to that shown in Refs.
\cite{Garcia-Ojalvo04,McMillen02,APikovskyBook01}. This pattern
indicates that a specific CRM would combine intracellular and
intercellular signals to coordinate the gene expressions in a
uniform way at the population level. Interestingly in the case of
ORN, we find that different initial conditions lead to three kinds
of dynamic patterns: 1-cluster, 2-cluster and 3-cluster
\cite{balancedcluster}. Similar phenomena were also found in a
chemical system \cite{Kiss05,AFTaylor08}. In the case of NOR,
however, neither synchronization nor clustering is observed, but an
interesting phenomenon that all cells are staggered equally in time,
i.e., so-called splay state, is found for the first time in a cell
population although the similar phenomenon was also detected
experimentally in a multimode laser system \cite{Splay}. Finally, in
the case of NAND, we also observe three types of clusterings at the
scattered initial states: 3-, 4- and 5-clusters
\cite{Ullner07,Golomb92}. The complete synchronization, however,
never occurs in this case.

To understand and interpret the above interesting patterns, we have
performed an analytical study of the system in the phase model
description \cite{BookKuramoto84}, which holds in a weak coupling
case. In this description, we first rewrite Eq. (2) as the following
symmetric form of coupling
\begin{eqnarray} \label{eq3}
\frac{dS_i}{dt}= E({\bf
X}_i,S_i)-\eta(1-Q)S_i+\frac{1}{N}\sum^{N}_{j=1}\eta Q(S_j-S_i)\,.
\end{eqnarray}
Then, for convenience the system consisting of both Eq. (1) and the
equation
\begin{eqnarray} \label{eq4}
\frac{dS_i}{dt}= E({\bf X}_i,S_i)-\eta(1-Q)S_i
\end{eqnarray}
is called as auxiliary system, which is assumed to generate a
sustained oscillation. For a weak coupling, the Kuramoto phase
reduction method \cite{BookKuramoto84} gives
\begin{eqnarray} \label{eq5}
\frac{d\phi_i}{dt}=
\omega_i+\frac{1}{N}\sum^{N}_{j=1}H_{ij}(\phi_j-\phi_i)\,,
\end{eqnarray}
where $\phi_i$ and $\omega_i$ stand for the phase and frequency of
the auxiliary system, respectively. $H_{ij}(\Delta \phi)$ represents
the interaction function with respect to the phase difference
$\Delta \phi=\phi_j-\phi_i$ between two cells,
\begin{eqnarray} \label{eq5}
H_{ij}(\phi_j-\phi_i)=\frac{1}{2\pi}\int_{0}^{2\pi}Z(\theta)\cdot
p(\phi_j-\phi_i+\theta)d\theta
\end{eqnarray}
which can be calculated numerically \cite{Ermentrout91}, where
$Z(\theta)$, a phase response function characterizing the phase
advance per unit perturbation, is a $2\pi$-period function, and
$p=(0,0,0,0,0,0,\eta Q(S_j-S_i))^{\rm T}$. Below we will omit
subscripts $i$ and $j$ for convenience. From $H(\Delta \phi)$, we
introduce a function: $G(\Delta \phi)= H(\Delta \phi)-H(-\Delta
\phi)$, to determine the mode of coupling. If $G(\Delta \phi)$
exhibits a positive slope at $\Delta \phi=0$, i.e, $G'(0)>0$, the
coupling is phase-attractive; If $G'(0)<0$, the coupling is
phase-repulsive. Therefore, Fig. 2 implies that the CRMs in the
cases of ANDN and ORN correspond to the phase-attractive coupling
whereas those in the cases of NOR and NAND correspond to the
phase-repulsive coupling. Such an approach based on the sign of
$G'(0)$ that depends generally on the intrinsic dynamics of the
uncoupled oscillator and on the interaction between the oscillators
is more effective than that of directly observing the network
topology in determining the mode of weak coupling \cite{Ullner07},
especially in the case of complex network architectures.

\begin{figure}[htb]
\begin{center}
\includegraphics [width=\hsize]{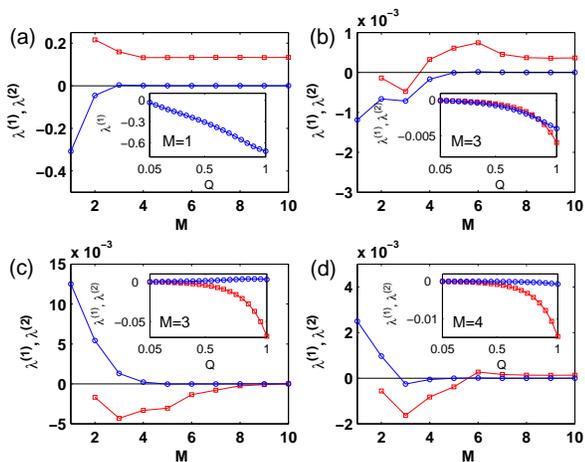}
\caption{(color). Eigenvalues associated with intra-cluster
fluctuations ($\lambda^{(1)}$: blue circle) and the maximal real
part of non-zero eigenvalues associated with inter-cluster
fluctuations ($\lambda^{(2)}$: red square) as a function of the
number of balanced clusters ($M$) for a fixed $Q=0.5$ in the case
of: (a) ANDN; (b) ORN; (c) NOR and (d) NAND. Insets: the dependence
relation of $\lambda^{(1)}$ and $\lambda^{(2)}$ on the parameter
($Q$) for a particular clustering as indicated.}\label{f1}
\end{center}
\vspace{-0.5cm}
\end{figure}

One cannot, however, obtain knowledge about clustering from the sign
of $G'(0)$. Since we are interested mainly in balanced clusters
\cite{balancedcluster}, we next employ Okuda's approach to determine
the stability of such clusters \cite{Okuda93}. In that method, we
need to calculate two kinds of eigenvalues: one is associated with
intra-cluster fluctuations and the other with inter-cluster
fluctuations, which are denoted by $\lambda_p$ and $\lambda_q$ (see
the caption of Fig. 3) respectively, where $M\leq p\leq N-1$ and
$0\leq q\leq M-1$ with $M$ being the number of clusters
presumptively. For convenience, denote by $\lambda^{(1)}$ and
$\lambda^{(2)}$ the $N-M$ same eigenvalues $\lambda_p$ and the
maximum of the the real parts of ($M-1$) non-zero eigenvalues
$\lambda_q$, respectively. Then, the stability of clusterings can be
determined by the signs of $\lambda^{(1)}$ and $\lambda^{(2)}$.
Specifically, the clustering is stable if both $\lambda^{(1)}$ and
$\lambda^{(2)}$ are negative, and unstable if $\lambda^{(2)}$ is
positive. In addition, if $\lambda^{(1)}$ is positive and
$\lambda^{(2)}$ is negative, and further if $M=N$, the $M$-cluster
(i.e., the splay state) are also stable. The dependence of
$\lambda^{(1)}$ and $\lambda^{(2)}$ on the balanced cluster number
$M$ in the cases of four CRIFs is shown in Fig. 3, which further
verifies the dynamic patterns shown in Fig. 2. The insets of Fig. 3
show that the parameter $Q$ has the significant influence on the
stability of clusterings (even including 1-cluster in Fig. 2(a) and
the splay state in Fig. 2(c)) for a particular balanced cluster
state, according to the above analysis.

In addition, in order to verify that the above results are of
generality, we also investigated the case of genetic relaxation
oscillators by using a detailed example studied in
Ref.\cite{McMillen02}, and found that different CRMs also drive
fundamentally different dynamic patters, but different types of
CRIFs would lead to different cellular patterns from those in the
case of repressilator (due to the paper length, the detailed results
are displayed in \cite{Supp}).

In summary, using models of synthetic genetic oscillators coupled to
quorum sensing, we have shown that different CRMs drive
fundamentally different cellular patterns, such as synchronization,
clustering, and splay state. Our results imply the following two
points: (1) Multicellular organisms possibly evolve into some
functional CRMs for particular goals by performing an elaborate
computation for input TFs; (2) Genetic network architecture found in
synchronous circadian clocks \cite{Garcia-Ojalvo04,JCDunlap99} might
be constrained since the complete synchronization independent of
initial conditions takes place only in the case of ANDN type of
responses. In particular, our results do suggest possible candidate
circuits for synchronous circadian clocks, while excluding others.
We expect that our theoretical findings will stimulate further
investigations under a more realistic condition involving
stochasticity \cite{JMRaser05,Zhou05} and spatial heterogeneousness
\cite{Basu05}, which would help us to understand differentiation
patterns and natural developmental processes.

We acknowledge the valuable comments and suggestions of anonymous
reviewers and the support from NSKF of P. R. of China (No.
60736028).


\begin{thebibliography}{27}
\bibitem{balancedcluster}
Here we consider balanced clusterings only. Other clusterings are
possible.
\end{thebibliography}
\end{document}